\documentclass[%
reprint,
superscriptaddress,
a4paper,
showkeys,
amsmath,amssymb,
aps,
pra,
]{revtex4-2}

\usepackage[english]{babel}
\usepackage[utf8]{inputenc}
\usepackage{amsmath,amsthm,amsfonts,amssymb,amscd,mathtools}
\usepackage[dvipsnames]{xcolor}
\usepackage[
    colorlinks,
    citecolor=RoyalBlue,
    linkcolor=RoyalBlue,
    urlcolor=RoyalBlue,
]{hyperref}
\usepackage{lipsum}
\usepackage{dcolumn}%
\usepackage{bm}
\usepackage{enumerate}
\usepackage{comment}
\usepackage{siunitx}

\usepackage{textcomp}
\usepackage{graphicx}
\usepackage{fixmath}
\usepackage{xcolor}
\usepackage[dvipsnames]{xcolor}

\newcommand{\boldsf}[1]{\textsf{\textbf{#1}}}

\usepackage{physics}

\newcommand{\qqoute}[1]{``#1''}

\newcommand{\iu}{\mathrm{i}\mkern1mu}

\usepackage{xspace}  
\usepackage{chemformula}
\newcommand{\VOtwo}{VO\textsubscript{2}\xspace}

\newcommand{\celsius}{\textcelsius\xspace}

\newcommand{\SiOtwo}{SiO\textsubscript{2}\xspace}
\newcommand{\SM}{Supplementary Material\xspace}

\def\beq{\begin{equation}}
\def\eeq{\end{equation}}

\newcommand{\cref}[1]{chapter~\ref{#1}}

\newcommand{\Cref}[1]{Chapter~\ref{#1}}

\def\3rd{3$^{\text{rd}}$}
\def\5th{5$^{\text{th}}$}

\begin{document}

\title{Switchable optical trapping of Mie-resonant phase-change nanoparticles}

\author{Libang Mao}
\altaffiliation{These authors contributed equally to this work}

\affiliation{Institute of Cardio-Cerebrovascular Medicine, Central Hospital of Dalian University of Technology, Dalian 116024, Liaoning Province, P. R. China} 
\affiliation{Nonlinear Physics Center, Research School of Physics, Australian National University, Canberra ACT 2601, Australia}
\affiliation{School of Optoelectronic Engineering and Instrumentation Science, Dalian University of Technology, Dalian, 116024, P. R. China} 

\author{Ivan Toftul}
\altaffiliation{These authors contributed equally to this work}
\affiliation{Nonlinear Physics Center, Research School of Physics, Australian National University, Canberra ACT 2601, Australia}

\author{Sivacarendran Balendhran}

\affiliation{Department of Electrical and Electronic Engineering, The University of Melbourne, Victoria, 3010 Australia}

\author{Mohammad Taha}
\email{Corresponding author: mohammad.taha@unimelb.edu.au}
\affiliation{Department of Electrical and Electronic Engineering, The University of Melbourne, Victoria, 3010 Australia}

\author{Yuri Kivshar}
\email{Corresponding author: yuri.kivshar@anu.edu.au}
\affiliation{Nonlinear Physics Center, Research School of Physics, Australian National University, Canberra ACT 2601, Australia}

\author{Sergey Kruk}
\email{Corresponding author: sergey.kruk@outlook.com}
\affiliation{Nonlinear Physics Center, Research School of Physics, Australian National University, Canberra ACT 2601, Australia}

\keywords{tunable optical force, phase-change material, Mie resonance, vanadium dioxide nanoparticles}

\begin{abstract}
Optical tweezers revolutionized the manipulation of nanoscale objects. Typically, tunable manipulations of optical tweezers rely on adjusting either the trapping laser beams or the optical environment surrounding the nanoparticles. We present a novel approach to achieve tunable and switchable trapping using nanoparticles made of a phase-change material (vanadium dioxide or \VOtwo). By varying the intensity of the trapping beam, we induce transitions of the \VOtwo between monoclinic and rutile phases. Depending on the nanoparticles’ sizes, they exhibit one of three behaviours: small nanoparticles (in our settings, radius  $<0.12$ wavelength $\lambda$) remain always attracted by the laser beam in both material phases, large nanoparticles ($>0.22 \lambda$) remain always repelled. However, within the size range of $0.12$--$0.22 \lambda$, the phase transition of the \VOtwo switches optical forces between attractive and repulsive, thereby pulling/pushing them towards/away from the beam centre. The effect is reversible, allowing the same particle to be attracted and repelled repeatedly. The phenomenon is governed by Mie resonances supported by the nanoparticle and their alterations during the phase transition of the \VOtwo. This work provides an alternative solution for dynamic optical tweezers and paves a way to new possibilities, including optical sorting, light-driven optomechanics and single-molecule biophysics.

\keywords{tunable optical force, phase-change material, Mie resonance, vanadium dioxide nanoparticles}
\end{abstract}

\maketitle

\section{Introduction and concept}
\label{sec:intro} 

Optical tweezers have become a powerful tool to precisely manipulate micro- and nanoparticles in a contactless way with widespread applications in biology, physics, and chemistry~\cite{choudhary2019bio,polimeno2018optical,hu2020near}. 
The trapped nanoparticle may be the object of study and manipulations, or it may often serve as an anchor point for the studied object, such as a single biomolecule bounded to its surface~\cite{Bustamante2021Mar}.
The basis of conventional optical tweezers is optical gradient forces that attract particles towards the intensity maximum of a tightly focused laser spot~\cite{nieminen2007physics}. Besides,
optical recoil forces, which are related to the directional scattering by the particle, extend the range of tweezer techniques, including 
optical pulling force~\cite{chen2011optical}, optical lateral force~\cite{Antognozzi2016Aug,wang2014lateral}, and orbiting motions of the particles~\cite{Tkachenko2020Jan,Simpson2010Sep}.

Compared to static optical forces, tunable optical forces have many potential applications in optical sorting~\cite{Wang2005Jan,Hoi2009Dec}, dynamic manipulation~\cite{ghosh2020next} and single-molecule biophysics~\cite{Bustamante2021Mar}. Generally, tunability can be introduced into optical tweezers in three conceptual ways: by tuning the laser, by adjusting the optical environment, or by modifying the particles themselves.

One of the more well-developed approaches is to tune the laser beams in free space~\cite{chen2023switchable}. 
A representative technique is the holographic optical tweezers, which usually use spatial light modulators (SLM) to dynamically control the optical force by adjusting the intensity, phase~\cite{nan2018sorting} and polarization distribution of light~\cite{nan2022creating,tang2021tunable}. 
While such systems enable flexibility, SLMs and their external controllers unavoidably increase their cost and size.

Second approach is to manipulate the optical environment in the vicinity of trapped nanoparticles with micro- and nanostructures such as metasurfaces~\cite{li2023switchable,geromel2023geometric,li2022experimental}, waveguides~\cite{zhang2017reconfigurable,Toftul2020Jan,Tkachenko2020Jan} or 2D material layers~\cite{cai2023tunable,paul2022tunable}. 
By adjusting the incident beam configuration (polarization~\cite{wang2018plasmonic,wang2011trapping}, wavelength~\cite{huang2018tunable,Kostina2019Mar}, focus position~\cite{messina2020two,ivinskaya2017plasmon}, etc.~\cite{huft2017holographic}) or changing the optical properties of constituent materials~\cite{zhang2017reconfigurable,cao2016controlling}, the near field or far field of these structures can be tuned dynamically, leading to the change of the resultant optical forces. For example, microscopic particles embedded with metasurface, called metavehicles, can be actuated by plane-wave illumination and steered by the polarization of the incident light~\cite{andren2021microscopic,qin2022tunable}. This approach can be more compact but requires delicate and expensive nanofabrication with top-down methods such as lithography techniques leading to challenges in the repeatability, and efficacy of optical manipulation techniques. 

In this study, we focus on the third approach, which is dynamically altering the properties of trapped particles to tune the optical forces, especially the Mie-resonant optical forces.  
Interestingly, the pioneers of optical trapping A. Ashkin and J. M. Dziedzic were discussing optomechanical manifestation of Mie resonances as early as in 1977~\cite{Ashkin1977Jun}. However, there was a long-lasting standstill in the research regarding Mie-resonant optomechanics until very recent years~\cite{Toftul2023Jun,kiselev2020multipole, Lepeshov2023Jun,Kislov2021Sep,Nan2023NC,Jiang2016Apr,Zhou2023Jun,Jiang2015Dec,Chen2014Sep,Kislov2021Sep}, including a theoretical model of how a specific Mie-resonant behaviour could potentially enhance the trapping performance~\cite{Lepeshov2023Jun}.
The interplay between Mie resonances has a profound effect on optical forces, for example, magneto-electric interference (Kerker interference)~\cite{xu2020kerker}, is quite different with the mechanism that governs conservative gradient forces and radiation forces~\cite{Ito2018May,Setoura2018Sep}, refreshing our understanding of optical forces~\cite{li2014ultrasensitive}. 
In particular, the sign and magnitude of these Mie-resonant optical forces can be theoretically tuned by changing the geometry parameters of the particles (shape and size)~\cite{Kislov2021Sep,Nan2023NC}. By changing the optical property of tunable materials, core-shell nanoparticles made of \ch{Ge2Sb2Te5} (GST)~\cite{cao2016fano} or black phosphorous (bP)~\cite{yang2018tunable} can also support tunable optical force theoretically. However, these designs remain difficult to implement in the experiment.

Here we employ nanoparticles made of a phase-change material  (\ch{VO2}) with their optical and optomechanical response governed by Mie resonances. We experimentally demonstrate the effect of optical force switching from attractive to repulsive through the manipulation of trapping laser power. The underlying mechanism is theoretically studied  by multipolar decomposition approach to illustrate the different force contributions and their origin. This work provides a feasible scheme to experimentally achieve tunable optical force, opening new avenues for dynamic manipulation and biophysics.

\section{\label{sec:experiment} Experimental results}

\begin{figure}
    \centering
    \includegraphics[width=\linewidth]{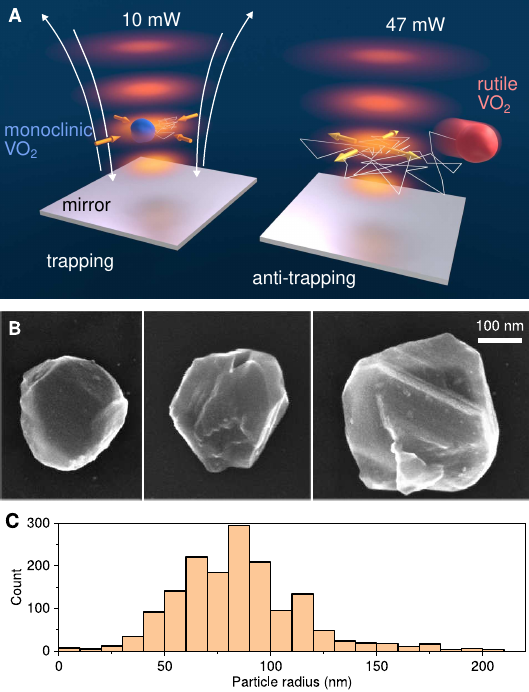}
    \caption{\textbf{Concept of switchable optical trapping.} 
    \boldsf{A} Optical forces acting on a \VOtwo nanoparticle switch from ``attractive" for the monoclinic (cold) state to ``repulsive" for the rutile (hot) state. 
    \boldsf{B} Scanning electron microscope images of the synthesized nanoparticles.
    \boldsf{C} Size distribution of the synthesized nanoparticles.}
    \label{fig:concept}
\end{figure}

We start our experiments from taking electron microscope images of the nanoparticles (Fig.~\ref{fig:concept}~\boldsf{B}). For this, we spin-coat a solution with nanoparticles onto a substrate and allow it to dry out. We mostly observe the nanoparticles with sizes (radii) varying from tens to hundreds of nanometres. We proceed with gathering statistical distribution of the nanoparticles' sizes in the solution. For this, we use a Nanoparticle Tracking Analysis instrument ZetaView, and the results are shown in Fig.~\ref{fig:concept}~\boldsf{C}.

We proceed with the study of individual particles under the optical force microscope. Supplementary Video 1 shows a representative example. We first record the Brownian motion of such particles with the trapping laser beam off in order to estimate their size using the tracking analysis method~\cite{Wagner2014May}. The nanoparticle size in Supplementary Video 1 is estimated to be $154\pm7$~nm. We proceed with optical trapping of the particle while gradually varying power of the trapping laser.
As illustrated in Fig.~\ref{fig:concept}, we employ a mirror-based geometry~\cite{Bowman2013Jan,Zemanek1999Nov,Zemanek1998Jun,Wu2017Feb} to balance the optical pressure in the axial direction (more details are in Methods). The phase transition of \ch{VO2} from monoclinic (we also refer to it as \qqoute{Cold \ch{VO2}}) to rutile (\qqoute{Hot \ch{VO2}}) is controlled optically by changing the intensity of the trapping beam. As shown in the Supplementary Material, those phases have  different refractive index values~\cite{Shao2018Jul}, which affects strongly the optical trapping. Our previous work shows that the resistance and transmittance of \ch{VO2} nanoparticles start to change at 36~\celsius and finish the phase transition at around 68 ~\celsius~\cite{Taha2023JMaterChemA}.
We observe that at lower laser powers, the particle can be trapped by an attractive optical force, and once the laser power exceeds its critical value, the nanoparticle escapes as the optical force switches from attractive to repulsive. The switching between attractive and repulsive forces is reversible and same particle can be attracted/repelled repeatedly. The Supplementary video 2 demonstrates clearly that the particle escapes the laser beam not due to the random Brownian motion, but due to a directional repulsive force (e.g. it is shot off from the beam center quickly).

\begin{figure*}
    \centering
    \includegraphics[width=0.99\linewidth]{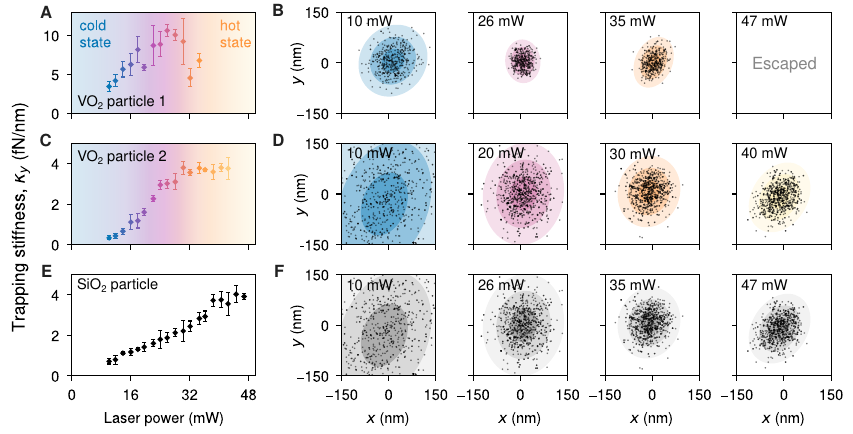}
    \caption{\textbf{Experimental observations}. \boldsf{A},\boldsf{C} Measured trapping stiffness of the two \VOtwo particles of different sizes as a function of the laser power. Past 26~mW the dependence of stiffness on power becomes strongly non-linear. For the particle 1 past 35~mW the optical trapping becomes repulsive and stiffness becomes undefined, while particle 2 still remains trapped. \boldsf{B},\boldsf{D} Measured particles displacement scatter plots at four levels of power.  Control experiment with a \SiOtwo particle under the same conditions. The \SiOtwo particle demonstrates linear dependence of stiffness vs power \boldsf{E} and monotonously decreasing displacement with increasing power \boldsf{F}.}
    \label{fig:experimental_observations}
\end{figure*}

We proceed with a quantitative study of this tunable trapping phenomenon experimentally by measuring the optical trapping stiffness of \VOtwo particles versus incident laser power. The trapping stiffness is estimated by calculating the variance in the Brownian motion of a trapped particle, describing how tight this particle can be trapped in the optical tweezers (see Methods). As can be seen in Fig.~\ref{fig:experimental_observations}~\boldsf{A}, the dependence of the trapping stiffness of the \VOtwo particle on the laser power deviates substantially from a monotonous linear trend, and after a threshold power, the particle becomes anti-trapped. We repeat out optical trapping experiments for several particles with a qualitatively similar example shown in the Supplementary Material. We perform a control experiment with \SiOtwo particles under the same conditions, and we observe in sharp contrast to \VOtwo a monotonous dependence of the stiffness versus laser power (see Fig.~\ref{fig:experimental_observations}~\boldsf{E} and~\boldsf{F}). 

The first region in Fig.~\ref{fig:experimental_observations}~\boldsf{A} is labeled as cold state with blue color. When the laser power is below $\sim 26$~mW, the trapping stiffness grows linearly with laser power and the ratio between trapping stiffness and laser power is consistent. Since this ratio is related to the refractive index mismatch between the particle and surroundings, it can be deduced that the refractive index of \VOtwo particles in this region is rather stable even though the temperature is increasing sustainably. This stability of refractive index implies that the local temperature on the \VOtwo particle is still below the phase transition threshold in this region. The second region is the intermediate state marked as magenta color. When the laser power is in the range 26--35~mW, the trapping stiffness deviates substantially from a linear trend. The third region (orange) represents the hot state. When the laser power goes beyond $\sim 35$~mW, the particle escapes from the laser focus, therefore the trapping stiffness can no longer be measured in this region.

Panels \boldsf{B}, \boldsf{D} and \boldsf{F} in Fig.~\ref{fig:experimental_observations} show the variation of the position distribution in \(xy\)-plane for the \VOtwo and \SiOtwo particles at different laser power. The particle position is tracked from each frame of the videos that record the Brownian motion of a trapped particle around the laser focus. The semitransparent ellipses represents the 1, 2, and 3 standard deviations with dispersions \(\expval{\sigma_{x,y}^2}\) being the semi-axis. The size of the ellipses reveals the relative magnitude of trapping stiffness, since the stiffness is inversely proportional to the standard deviation, \(\kappa_{x,y} \propto 1/\expval{\sigma_{x,y}^2}\) (see Methods). It can be seen that the size of the color region first reduces and then enlarges when the laser power changes from 10~mW to 35~mW. When the laser power is $\sim 47$~mW, the particle will just escape, so there is no position distribution for this case.

We generally note that the effect of switching depends not only on the material (\ch{VO2}) but it also depends crucially on the particle size. Fig.~\ref{fig:experimental_observations}~\boldsf{C} and~\boldsf{D}  shows that a different size \ch{VO2} particle remains always trapped within the range of our laser power sweep. In accord with our theory, the phase transition in this particle changes the stiffness vs power dependence, but does not switch the attractive optical force to a repulsive force. Supplementary video 3 shows an example of an always repelled particle. The particle, in agreement with our theory, is being repelled by the laser beam even at low excitation powers. The particle radius is estimated to be $334\pm6$~nm, which is coincident with our theoretical prediction that the particle larger than 230 nm will be always repelled.

To verify that the variation trends in Fig.~\ref{fig:experimental_observations}~\boldsf{A} and Fig.~\ref{fig:experimental_observations}~\boldsf{B} are related to the \VOtwo phase transition, we conducted a control experiment using silica particles ($R \approx 1030~\text{nm}$) trapped under the same conditions. Fig.~\ref{fig:experimental_observations}~\boldsf{E} shows that the trapping stiffness of the silica particle has a trivial linear uptrend, which is in striking difference to the trend in Fig.~\ref{fig:experimental_observations}~\boldsf{A}. Fig.~\ref{fig:experimental_observations}~\boldsf{F} further visualizes the density change of the position distribution for the silica particle under four different laser power. The size of the standard deviation ellipse continues to decrease when the laser power increases from 10~mW to 47~mW, which is different with Fig.~\ref{fig:experimental_observations}~\boldsf{B}, demonstrating the difference with  the \VOtwo particle with phase transition.

To demonstrate that the particle escaping cannot be caused by the thermal diffusion, the impact of the Brownian fluctuation on the trapping stability is discussed in the following. 
For the reasonable powers the ratio $U_{\text{tr}}/(k_{\text{B}}T)$ remain constant as both trapping potential $U_{\text{tr}}$ and thermal energy $k_{\text{B}}T$ depend linearly on the intensity. For example,
it was experimentally verified by Perkins et al \cite{Seol2006Aug} that for absorptive gold nanoparticles, the trapping stiffness dependence on the laser power was linear, even though the temperature of the particle increased from 20~\celsius to around 75~\celsius. Moreover, the temperature dependence of trapping stiffness at a fixed laser power was also experimentally studied by Daniel et al \cite{Lu2020Nov}. In this reference, the trapping stiffness had a linear decrease of only 0.35 fN/nm when the temperature of the chamber changed from 20~\celsius to 68~\celsius. The effective radius of the nanoparticle was 120~nm, which is comparable to the \VOtwo particles in our manuscript. The Brownian motion dynamic simulation by Patricia et al \cite{Lu2021Aug} also showed that for a \SiOtwo microparticle, the trap stiffness at a fixed laser power only changed from 9.3 fN/nm to 8.4 fN/nm and the probability of escaping remained almost zero when the temperature of the environment increased from 20~\celsius to 80~\celsius (which in relation to our work would already be past the boiling point of ethanol). 
Furthermore, we believe that the temperature variation on the gold mirror is negligible in our setting because the laser focus in our experiment is several micrometers away from the gold.
To demonstrate this, we have simulated the temperature distribution on the gold mirror by the COMSOL. As can be seen in the Supplementary Figure 3, because of the heat transfer, the temperature on the gold mirror is very close to the room temperature, which will not cause any significant thermal effect. 

\section{\label{sec:theory} Theoretical approach}

To gain our understanding of the size-dependant behavior of the nanoparticles, we proceed with theoretical analysis.
Our analysis is based on the multipolar decomposition approach being a versatile tool for analysing the interaction between light and resonant nanoparticles~\cite{Gladyshev2020Aug,Poleva2023Jan,Alaee2019Jan}.
This technique relies on the multipole expansion of the scattered electromagnetic fields, allowing for a detailed and nuanced understanding of the light scattering, internal polarization current  distributions, absorption, and the resultant forces on nanoparticles. 
By decomposing these interactions into simpler multipolar components --- such as dipole and quadrupole --- this approach provides a thorough analysis of underlying optical phenomena. 
This is especially relevant for particles with complex material characteristics such \VOtwo nanoparticles, which have changing stimuli-depended properties.

\begin{figure}
    \centering
    \includegraphics[width=1\linewidth]{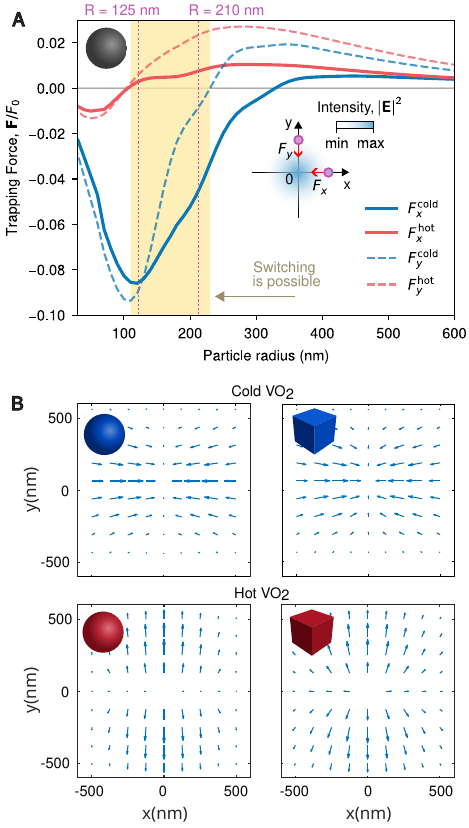}
    \caption{
        \textbf{Theoretical analysis}. (\boldsf{A}) Exact calculations of an optical force acting on the spherical \VOtwo nanoparticles in ethanol as  function of the particles' radius.
        The numerical aperture of the lens is set to the  \(\mathrm{NA} =0.9\). 
        The operation wavelength is \(\lambda = 980~\text{nm}\).
        \VOtwo particles in their cold state have a region of radii experiencing negative (attractive) optical force. Hot \VOtwo particles experience positive (repelling) optical force over the broad range of studied radii. Orange section shows the narrow region of particle radii for which switching from trapping to anti-trapping is possible. Inset shows the positions of the particles for the $F_x$ and $F_y$ force calculations. Distance from the beam center was chosen to be 250~nm. The polarization of the beam is aligned with the $x$-axis. 
        (\boldsf{B}) Comparison of spatial distributions of optical forces for  spherical and cubic particles of the equivalent volume. The radius of the sphere is 210~nm. The distance between the particle and gold mirror is set to be 3000 nm.
    }
    \label{fig:ott_results}
\end{figure}

At the first step we calculated the \textit{exact} optical force on spherical \VOtwo particles using optical tweezer toolbox~\cite{OTT,OTT_Nieminen2007Jul} which shown in Fig.~\ref{fig:ott_results}. 
This toolbox performs the calculations based on the transfer matrix theory.
We consider a non-paraxial highly focused Gaussian beam with \(\mathrm{NA} = 0.9\). 
In our model nanoparticles are placed in the focal plane with a constant lateral offset of \(250~\text{nm}\) from the beam center as shown in the inset of the Fig.~\ref{fig:ott_results}.
We calculate the dependence of the $F_x$ for the offset along the  $x$-axis and $F_y$ for the offset along the $y$-axis for both hot and cold states.
In calculations, two states differed in particles permittivity: $\varepsilon_{\text{VO}_2}^{\text{cold}} = 6.55+\iu 3.66$ and $\varepsilon_{\text{VO}_2}^{\text{hot}} = 0.68 + \iu 3.40$ at the operation wavelength $\lambda = 980~\text{nm}$. 
Particles were suspended in ethanol with $\varepsilon = 1.85$~\cite{Rheims1997Jun}. 
The \VOtwo dispersion was taken from the~\cite{Wan2019Oct} and is also plotted in the Supplementary Material.
The alteration of the refractive index of \VOtwo particles from cold state to hot state, leads to the fact that the force can change from negative to positive when the radius is between 120~nm and 230~nm (orange shaded region in Fig.~\ref{fig:ott_results}~\boldsf{A}), which means in this range \textit{the force switching is possible} --- the optical force is attractive for the cold state and repulsive for the hot state. 
In other words, the particle can be trapped or repelled by the laser focus depending on the trapping laser power.
We compare the results of our calculations for several examples of spherical and non-spherical particles of same volumes, and observe similar behavior (see Fig.~\ref{fig:ott_results}~\boldsf{B} and Supplementary Material). 
This is due to the nature of the Mie resonances which are mainly determined by the volume rather then the exact shape.
We thus focus on spherical geometries of the nanoparticles in our theory as a good approximation for the particles in our experiments (see Fig.~\ref{fig:concept}~\boldsf{B}). The total force distribution of \VOtwo particles with different shapes (sphere, hexagonal prism and cube) is also simulated by Ansys Lumerical FDTD (see Supplementary Material), showing that the switching optical trapping is also possible for irregular particles. Additionally, for comparison, we perform theoretical calculations for the \SiOtwo particles (presented in the Supplementary Material). As expected, \SiOtwo particles do not show switchable behavior.

Next, in order to explain what exactly causes the switching from trapping to anti-trapping we use the multipolar decomposition approach. 
Once we assume particles to have spherical shape, its optical linear response is explained fully analytically by the Mie theory~\cite{Gouesbet_generalized_mie_book,bohren2008absorption}. 
The scattered field can be decomposed into to the vector spherical harmonics and corresponding excitation strength, i.e. polarizabilities, are proportional to the electric $a_n$ and magnetic $b_n$ Mie scattering coefficients, where $n$ is the multipolar order.
\begin{figure*}
    \centering
    \includegraphics[width=0.99\linewidth]{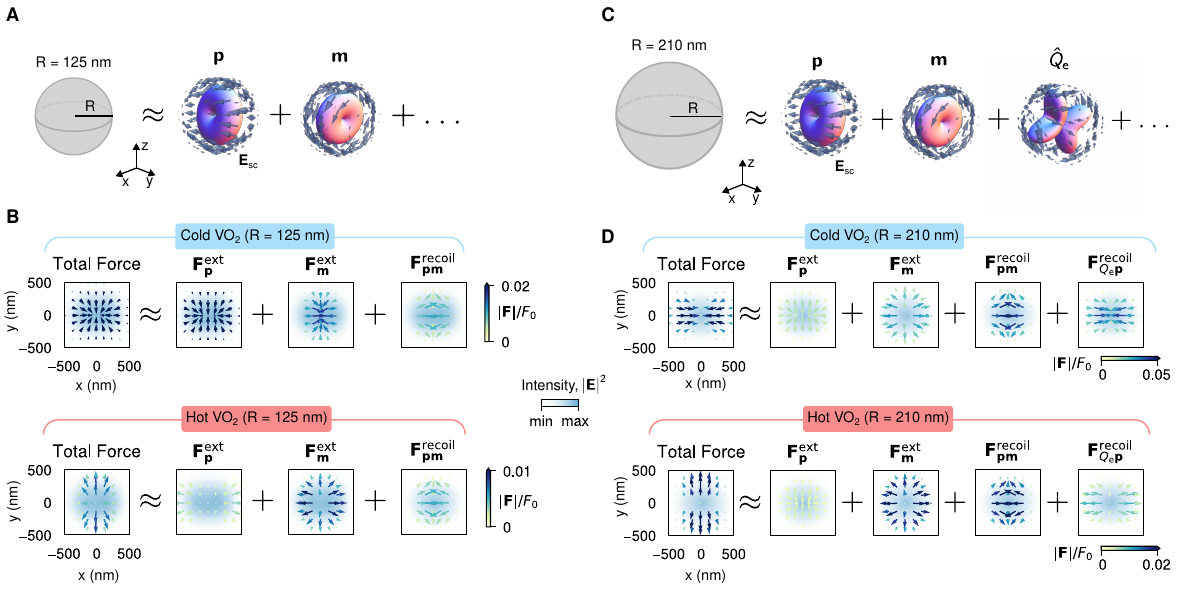}
    \caption{\textbf{Optical force multipolar decomposition}. \boldsf{A} \VOtwo particle of 125~nm radius can be approximately described by the electric and magnetic dipoles. Solid surface shows \(\abs{\vb{E}_{\text{sc}}}^2\) in the far field(See Supplementary Material), and arrows show \(\Re \left( \vb{E}_{\text{sc}} \right)\). The excited electric dipole has a non-zero $p_z$ component which results in a tilted donut shape.
    \boldsf{B} Multipolar decomposition of the transverse optical force $(F_x, F_y)$ for the \textit{smaller} particle size in the switching region. Dominant force contributions are the electric dipole force and magnetic dipole force, while dipole-dipole recoil contribution $\vb{F}^{\text{recoil}}_{\mathbf{p} \mathbf{m}}$ brings the asymmetry in the final force distribution.
    \boldsf{C} \VOtwo particle of 210~nm radius can be approximately described by the electric dipole, magnetic dipole, and electric quadrupole. 
    \boldsf{D} Multipolar decomposition of the transverse optical force $(F_x, F_y)$ for the \textit{bigger} particle size in the switching region. As the particle becomes bigger, the higher multipoles start to play a significant role. In particular, the recoil force based on the electric dipole-quadrupole interference $\vb{F}^{\text{recoil}}_{Q_{\text{e}} \vb{p}}$ helps switching mechanism for particle radius of $R = 210~\text{nm}$.
    All components for the bigger particle are plotted in the \SM.}
    \label{fig:force_multipole_decomposition}
\end{figure*}

The time average force for the monochromatic case is equal to the average rate of linear momentum change. 
This can be written in terms of the surface integration of the linear momentum flux as \(\vb{F} = - \oint \limits_\Sigma \tensor{\mathcal{T}} \cdot \dd \vb{\Sigma}\), where integration is performed over a surface enclosing the particle \(\Sigma\) and \(\tensor{\mathcal{T}}  = -\frac{1}{2} \Re \left[  \varepsilon \varepsilon_0 \vb{E}^* \vb{E} + \mu \mu_0 \vb{H}^* \vb{H} - \frac{\tensor{I}}{2} \left( \varepsilon \varepsilon_0 \abs{\vb{E}}^2 + \mu \mu_0\abs{\vb{H}}^2\right)\right]\) is the linear momentum flux of the electromagnetic field (or minus Maxwell stress tensor)~\cite{novotny2012principles,Shi2022Sep}.
From here is follows that the optical force is a quadratic function of the fields or linear function of the intensity. 
During scattering process total field can be written as a sum of incident and scattered ones as \(\{\vb{E},\vb{H}\} = \{\vb{E}_{\text{inc}},\vb{H}_{\text{inc}}\} + \{\vb{E}_{\text{sc}},\vb{H}_{\text{sc}}\}\), and hence \(\tensor{\mathcal{T}} = \tensor{\mathcal{T}}_{\text{inc}} + \tensor{\mathcal{T}}_{\text{mix}} + \tensor{\mathcal{T}}_{\text{sc}}\). 
Since incident field flux $\tensor{\mathcal{T}}_{\text{inc}}$ does not contribute to the change of linear momentum in a lossless medium, the optical force can be decomposed into two main terms
\begin{equation}
    \vb{F} = \vb{F}_{\text{mix}} + \vb{F}_{\text{recoil}},
\end{equation}
where $\vb{F}_{\text{mix}} = - \oint \limits_\Sigma \tensor{\mathcal{T}}_{\text{mix}} \cdot \dd \vb{\Sigma}$ is the  part of the force which emerges from the interference between incident field and scattered field; $\vb{F}_{\text{recoil}} =-  \oint \limits_\Sigma \tensor{\mathcal{T}}_{\text{sc}} \cdot \dd \vb{\Sigma}$ is the recoil part of the force which is related to the directional scattering of the particle~\cite{Salandrino2012Apr,Duan2017Nov,Chen2014Sep,Nieto-Vesperinas2010May}. 

We can apply multipolar expansion of the scattered field only~\cite{Chen2014Sep,Jiang2016Apr,Zhou2023Jun,Jiang2015Dec,Chen2014Sep,Kislov2021Sep}.
Since all the multipole polarizabilities are intricately linked with the Mie scattering coefficients, it becomes possible to estimate the necessary number of multipoles to be taken into consideration based on the size parameter \(kR\).
This can be achieved by examining the magnitudes of \(\abs{a_{n}}\) and \(\abs{b_{n}}\) (see \SM). 
Up to a quadruple moments optical force can be written as
\begin{align}
    \vb{F}_{\text{mix}} &= \vb{F}^{\text{mix}}_{\vb{p}} + \vb{F}^{\text{mix}}_{\vb{m}}  + \vb{F}^{\text{mix}}_{Q_{\text{e}}} + \vb{F}^{\text{mix}}_{Q_{\text{m}}} + \dots \\
    \vb{F}_{\text{recoil}} &= \vb{F}^{\text{recoil}}_{\vb{p} \vb{m}} + \vb{F}^{\text{recoil}}_{Q_{\text{e}} \vb{p}} + \vb{F}^{\text{recoil}}_{Q_{\text{m}} \vb{m}} + \vb{F}^{\text{recoil}}_{Q_{\text{e}} Q_{\text{m}}} + \dots
\end{align}
where index represent from which multipole this force arose from. Here $\vb{p}$ and $\vb{m}$ are the electric and magnetic dipoles, and $Q_{\text{e}}$ and $Q_{\text{m}}$ are the electric and magnetic quadrupoles. If there are two indices, this means that this component of the forces has the origin from the interference pattern of the corresponding multipoles. In the region of interest, i.e. for the \VOtwo particles within the $120-230~\text{nm}$ range, the main contribution to the total force are from the electric and magnetic dipoles, as well as two recoil contributions from dipole-dipole and dipole-quadrupole interference. We write these contributions explicitly 
\begin{align}
    \vb{F}^{\text{mix}}_{\vb{p}} &= \frac{1}{2} \Re \left[ \vb{p}^*  \cdot (\grad) \vb{E}\right], \\
    \vb{F}^{\text{mix}}_{\vb{m}} &= \frac{\mu \mu_0}{2} \Re \left[ \vb{m}^*  \cdot (\grad) \vb{H}\right], \\
    \vb{F}^{\text{recoil}}_{\vb{p} \vb{m}}  &= -\frac{k^4}{12 \pi} \sqrt{\frac{\mu \mu_0 }{\varepsilon \varepsilon_0}} \Re( \vb{p}^* \cp \vb{m}), \\
    \vb{F}^{\text{recoil}}_{Q_{\text{e}} \vb{p}} &= - \frac{k^5}{40 \pi \varepsilon \varepsilon_0} \Im \left[ \hat{Q}_{\text{e}} \vb{p}^* \right], 
\end{align}
where \(\vb{p} = \varepsilon \varepsilon_0 \alpha_{\text{e}} \vb{E}\) and \(\vb{m} = \alpha_{\text{m}} \vb{H}\)  are the induced electric and magnetic dipole moments, \(Q_{\text{e},ij} = \varepsilon \varepsilon_0 \alpha_{Q_{\text{e}}} \left( \nabla_i E_j + \nabla_j E_i \right)/2\) is the induced electric quadrupole moment. Polarizabilities are related with the Mie coefficients as \(\alpha_{\text{e}} = \iu 6\pi k^{-3} a_1\), \(\alpha_{\text{m}} = \iu 6\pi k^{-3} b_1\), and \(\alpha_{Q_{\text{e}}} = \iu 40\pi k^{-5} a_2\). The explicit form of the Mie coefficients can be found in Ref.~\cite{bohren2008absorption}. The general form for the force and polarizabilities in all orders can be found in Ref.~\cite{Jiang2016Apr}. The \qqoute{\(\cdot (\grad)\)} is so-called Berry notation and should be read as \(\vb{A}\cdot (\grad) \vb{B} = \sum \limits_{j} \vu{e}_j \sum \limits_{i = x,y,z} A_i \nabla_j B_i\) for arbitrary vectors \(\vb{A}\) and \(\vb{B}\).

\begin{figure*}
    \centering
    \includegraphics[width=0.9\linewidth]{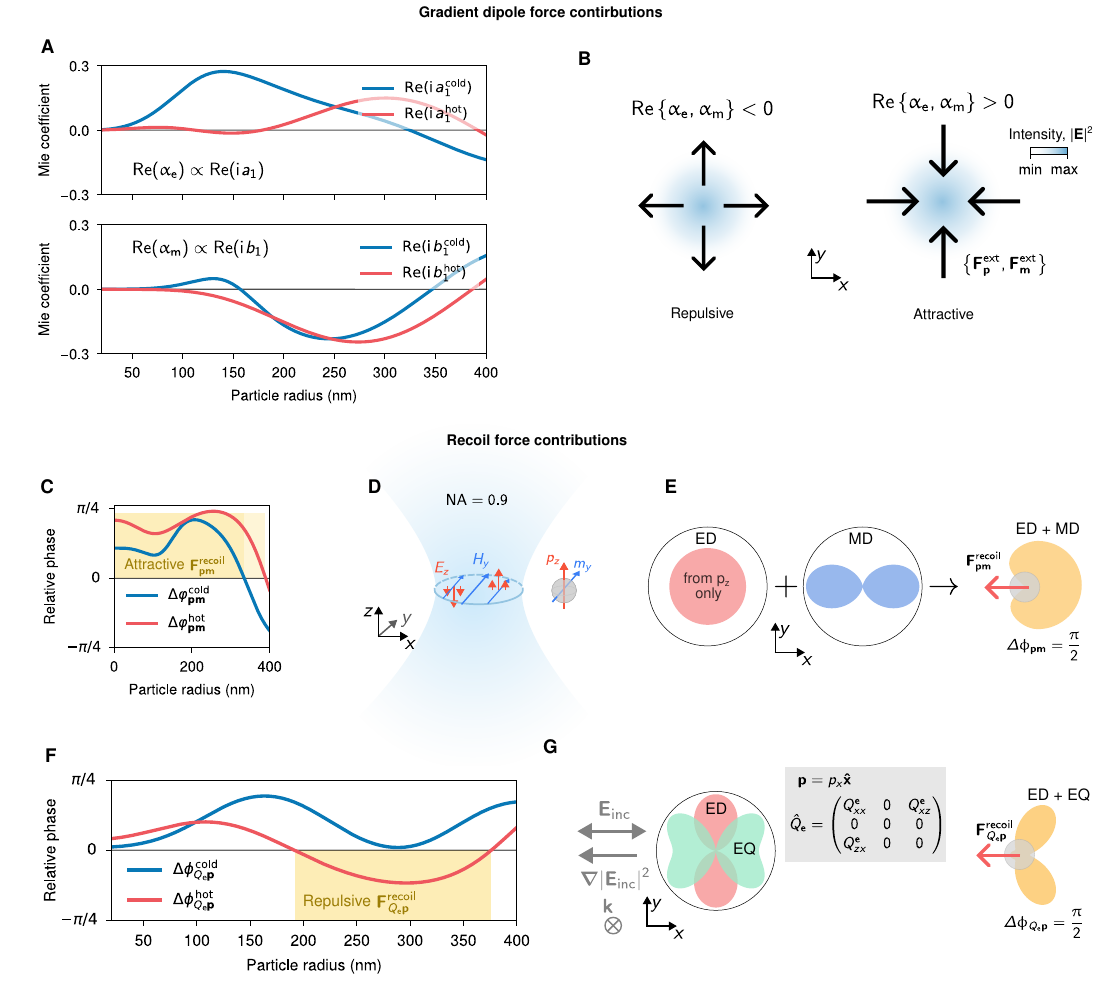}
    \caption{
        \textbf{Gradient dipole and recoil forces in-depth analyses}.  Electric and magnetic dipole polarizabilities are directly connected with the Mie scattering coefficients. Plot \boldsf{A} shows the sign of the real part of the $\alpha_{\text{e}}$ and $\alpha_{\text{m}}$ polarizabilities. \boldsf{B} Once $\Re \left\{ \alpha_{\text{e}}, \alpha_{\text{m}} \right\}$ is positive then the corresponding gradient part of the dipole force is going to be attractive; once negative --- repulsive. Panels \boldsf{C},\boldsf{D},\boldsf{E} and \boldsf{C},\boldsf{G}  show the recoil force contributions $\vb{F}^{\text{recoil}}_{\vb{p} \vb{m}}$ and $\vb{F}^{\text{recoil}}_{Q_{\text{e}} \vb{p}}$, correspondingly. 
    }
    \label{fig:phase_story}
\end{figure*}

Before analyzing the optical force switching region, we show that in the Rayleigh limit the switching is impossible within our accessible parameter range, and the optical force is always attractive no matter which phase of the \VOtwo we have. 
Deeply sub-wavelength sized particle (\(kR \ll 1\)) can be fully described by the dipole moment only \(\vb{p} = \varepsilon \varepsilon_0 \alpha_{\text{e}} \vb{E} \), where electric dipole polarizability is \(\alpha_{\text{e}}\approx 4\pi R^3 \frac{\varepsilon_{\text{p}} - \varepsilon}{\varepsilon_{\text{p}} +2 \varepsilon}\).
The gradient part of the dipole force simplifies to $\vb{F}_{\vb{p}} \approx 0.5 \Re\alpha_{\text{e}} \grad \abs{\vb{E}}^2$ (here we neglect the pressure force, i.e. contributions from proportional to $\Im \alpha_{\text{e}}$ as this does not contribute to the transverse trapping). 
For both cold and hot states at the operation wavelength $\lambda = 980~\text{nm}$ the real part of the polarizability is positive, i.e. \(\Re \alpha_{\text{e}} (\varepsilon^{\text{cold}}_{\text{\VOtwo}}) > 0\) and \(\Re \alpha_{\text{e}} (\varepsilon^{\text{hot}}_{\text{\VOtwo}}) > 0\). 
It implies that \(\vb{F}_{\text{p}}\) in Rayleigh limit is always attractive and optical force switching is not possible. 
However, for the bigger particles electric dipole contribution to the force, as well as magnetic dipole, can change its sign as illustrated in Fig.~\ref{fig:phase_story}~\boldsf{A} and~\boldsf{B}.

Next we analyze the switching region. First, we consider smaller particle of the size \(R = 125~\text{nm}\) at the left side of the region (Fig.~\ref{fig:force_multipole_decomposition}~\boldsf{A} and~\boldsf{B}). 
The Rayleigh limit is no longer applicable, and the presence of the magnetic dipole plays a significant role. 
We observe that in the cold state both electric dipole and magnetic dipole forces are attractive, however, in the hot state the electric dipole force is suppressed, while the \textit{magnetic dipole force becomes repulsive}. This correlates very nicely with the polarizability analyses in Fig.~\ref{fig:phase_story}~\boldsf{A}. As one can see in Fig.~\ref{fig:ott_results}~\boldsf{B}, the total force in hot and cold states is slightly asymmetric. This is due to the first recoil term \(\vb{F}^{\text{recoil}}_{\vb{p}\vb{m}}\), which is related to the directional scattering of the particle in the transverse plane. 
In paraxial approximation for which $E_z =0$ there is no transverse components of the \(\vb{F}^{\text{recoil}}_{\vb{p}\vb{m}}\) as dipole Kerker effect can be only forward or backward. In our setup we have used the objective lens with a high numerical aperture \(\mathrm{NA} = 0.9\). This brings significant non-zero longitudinal field components (up to a \(\operatorname{max}\abs{E_z} \approx 0.4 \operatorname{max}\abs{E_x}\), see also \SM). 
As shown in Fig.~\ref{fig:phase_story}~\boldsf{D}, for the highly focused \(x\)-polarized beam particle displaced from the center along the \(x\)-axis has both \(p_z \propto E_z\) and \(m_y \propto H_y\) which brings non-zero \({F}^{\text{recoil}}_{\vb{p}\vb{m},x} \neq 0\) due to the directional scattering in the \(xy\)-plane (Fig.~\ref{fig:phase_story}~\boldsf{E}). 
For the particle radii of up to almost \(R \lesssim 350~\text{nm} \) its contribution is attractive but not repulsive. This can be understood by looking at the relative phase between the electric and magnetic dipoles \(\Delta \varphi_{\vb{p} \vb{m}} = \arg a_1 - \arg b_1\), which is plotted in Fig.~\ref{fig:phase_story}~\boldsf{C}.

Secondly, we discuss the particle at the right edge of the switching region, in particular \(R = 210~\text{nm}\). Compared to the smaller size, now the electric quadrupole also plays an important role (Fig.~\ref{fig:force_multipole_decomposition}~\boldsf{C}).
The main contributions to the optical force are shown in Fig.~\ref{fig:force_multipole_decomposition}~\boldsf{D}. 
The most noticeable difference is the significance of the \(\vb{F}^{\text{recoil}}_{Q_{\text{e}} \vb{p}}\) which takes its origin in the electric dipole-quadrupole inteference and corresponding directional scattering, which illustrated in the Fig.~\ref{fig:phase_story}~\boldsf{G}.
The relative phase between the electric dipole and electric quadrupole determines whether this contribution would be repulsive or attractive. 
From the analyses of this phase \(\Delta \varphi_{Q_{\text{e}} \vb{p}} = \arg a_2 - \arg a_1\) in Fig.~\ref{fig:phase_story}~\boldsf{F}, we observe that for hot particle this contribution becomes repulsive, which supports the particle escape.

\section{\label{sec:conclusions} Conclusions}

In conclusion, we experimentally demonstrated switchable trapping and anti-trapping events using vanadium dioxide nanoparticles. The behaviour is driven by the phase transition of the \VOtwo induced all-optically by varying the intensity of the laser beam. The effect is governed by Mie multipoles supported by the particles and the changes in the multipolar composition occurring with the phase transition. Our theoretical analysis shows that the switchable behaviour is possible within a finite range of particles' sizes, which can be use for particle size sorting. On the lower end of the particle size, the effect is dominated by the electric and the magnetic dipoles. On the higher end of the particle size, the critical contribution comes from the recoil electric quadrupole. 

This work incorporates the Mie-resonant phase-change nanoparticles into optical tweezers technology, which brings the ability to manipulate optical forces between attractive and repulsive. With biomolecules bonded to the outer surfaces, the phase-change nanoparticles can be utilized as a carrier~\cite{cao2016fano} to dynamically sort the molecules on the surface. Besides, in most of single-molecule experiments~\cite{Bustamante2021Mar}, a single molecule is attached to the surface of polystyrene or silica particles, acting as a handle to manipulate and study single molecules by optical force. By replacing the static handle with the phase-change nanoparticles, optical tweezers may bring new possibilities for single molecule biophysics. Furthermore, the phase-change nanoparticles can be modified to have a transition temperature as low as 30~\celsius by inducing strain in the material~\cite{Taha2023JMaterChemA}. So the nanoparticles can also act as an all-optical power limiter at the nanoscale: the nanoparticle leaves the tweezer once critical power level is exceeded to protect the biomolecules and cells from heating damage. This novel approach, extends the capabilities of optical tweezers beyond traditional static methods, opening new avenues for a new generation of active optomechanical systems.
 
\section*{Methods}

\textbf{Optical tweezer setup}.
A NIR fiber-coupled laser (Thorlabs BL976-PAG500, 976 nm) coupled to an objective lens (Olympus MPlanFLN 100X, $\mathrm{NA} = 0.9$) was used as the trapping beam. A high-resolution CMOS camera (Thorlabs DCC3260M) was used to capture the images and record the videos of the optically trapped particles. The data detecting frame rate of the CMOS is 71 Hz, with detecting area of 160 pixels by 142 pixels. The experiment is performed at the room temperature (20~\celsius). The laser power is measured after the transmission of the objective. To balance the force on z direction, 100 nm gold was deposited by E-beam evaporation on the glass as the top of the chamber. 
The non-functionalized fused silica beads  (P/N SS04N/9857, diameter 2.06 \textmu m) are from Bangs Laboratories, Inc.

\textbf{Synthesis of nanoparticles}.
Vanadium pentoxide (\ch{V2O5}) and oxalic acid dihydrate in a hyperhydrated environment of reflux reactions are used here to synthesize \VOtwo nanoparticles~\cite{Taha2023JMaterChemA}. Breifly, the mixture is heated to 240~\celsius  and stirred at a high speed of 450 rpm, ensuring uniform heat distribution and reactant mixing. 
The elevated temperature is critical for initiating the chemical reactions, while the stirring speed helps in maintaining the consistency of the mixture. The process undergoes a reflux reaction for 48 hours, which maintains a constant temperature environment of approximately 100~\celsius. This temperature control is crucial for the reaction, as it allows for controlled kinetics and consistent product quality, without the need for the conventional hydrothermal autoclave processes typically used in nanoparticle synthesis. 
Post-reaction, the products are washed, centrifuged, filtered  and dried in ambient environment to remove impurities and excess water. The end product used in this study is sub-stoichiometric and hydrated \VOtwo nanoparticles. Detailed material characterization and compositional analyses can be found in our previous work \cite{Taha2023JMaterChemA}. The synthesis process flow use here is outlined in Supplementary Material.
Compared to traditional methods, this benchtop process offers a simpler, less energy-intensive approach, and reduces the risk of over-hydration of the metal oxides. 

\textbf{Trap stiffness detection method}.
We employ a video tracking analysis method to determine the 2D trap stiffness of optically trapped particles. Assuming linearity of the optical force near the trapping point $\vb{r}_0$, i.e., $\vb{F}(\vb{r}) \approx - \kappa (\vb{r} - \vb{r}_0)$, where $\kappa$ denotes the trap stiffness. According to the \textit{equipartition theorem}, there is $k_{\text{B}} T /2$ of thermal energy for each degree of freedom, where $T$ is the absolute temperature and $k_{\text{B}}$ is the Boltzmann constant. For the $x$-motion of the particle, we express the average potential energy of the trapped particle as $\kappa_x \langle\left( x - x_0 \right)^2 \rangle /2 = k_{\text{B}}T / 2$.
This implies that the trap stiffness can be computed by observing the Brownian motion of the particle and consequently determining the position variance $\expval{\sigma_{x}^2} = \expval*{\left( x - x_0 \right)^2}$. Thus, $\kappa_{x,y} =  k_{\text{B}} T/\expval{\sigma_{x,y}^2}$, where $T$ represents the operating temperature in Kelvin.
The 2D position of an optically trapped particle is obtained by localizing the center of the nanoparticle.
The recorded video was split into 200 frames fragments, and trap stiffness is calculated for the each segment. Using this data the standard deviation of trapping stiffness were calculated which is presented as error bars.

\section*{Acknowledgements}
Authors thank Dmitry Pidgayko,  Cheng-Wei Qiu, Yuzhi Shi, and Mikhail Petrov for a constructive criticism and useful suggestions. 
This work was supported by the Australian Research Council (grants DE210100679 and DP210101292) and Dalian University of Technology. Vanadium oxide nanoparticle research was supported by the Defence Science Institute, an initiative of the State Government of Victoria. This work used the ACT node of the NCRIS-enabled Australian National Fabrication Facility (ANFF-ACT). The authors acknowledge the help and support provided by the Cytometry team ( Harpreet Vohra and Michael Devoy) from the CHASM facility within JCSMR.

\bibliography{bibliography_VO2}

\end{document}